\documentclass[aps,twocolumn,showpacs,superscriptaddress,groupedaddress]{revtex4} 
\usepackage{graphicx,amssymb,epsfig,amsmath}

\usepackage{dcolumn}   
\usepackage{bm}        
\usepackage[loose]{units}

\begin{document}
\newcommand{\mv}{\mathbf{M}} 
\newcommand{\Hv}{\mathbf{H}} 
\newcommand{\hv}{\mathbf{h}} 
\newcommand{\ev}{\hat{\mathbf{e}}} 
\newcommand{\largesample}{$100 \times100 \; \unit{nm}^2$ sample } 
\newcommand{\smallsample}{$50  \times 50 \; \unit{nm}^2 $ sample }

\title{Spin-transfer pulse switching: From the dynamic to the thermally activated regime}

\author{D.  Bedau}
\author{H. Liu}
\affiliation{Department of Physics, New York University, New York, NY 10003 USA}
\author{J. Z. Sun}
\affiliation{IBM T. J. Watson Research Center, P.O. Box 218, Yorktown Heights, New York, 10598 USA}
\author{J. A. Katine}
\affiliation{San Jose Research Center, Hitachi-GST, San Jose, California 95135 USA}
\author{E. E. Fullerton}
\affiliation{CMRR, University of California, San Diego, La Jolla, California 92093 USA}
\author{S. Mangin}
\affiliation{IJL, Nancy-Universit{\'e}, UMR CNRS 7198, F-54042 Vandoeuvre Cedex, France}
\author{A. D. Kent}
\affiliation{Department of Physics, New York University, New York, NY 10003 USA}

\date{September 27, 2010}

\begin{abstract}
The effect of thermal fluctuations on spin-transfer switching has been studied for a broad range of time scales (sub-ns to seconds) in a model system, a uniaxial thin film nanomagnet. The nanomagnet is incorporated into a spin-valve nanopillar, which is subject to spin-polarized current pulses of variable amplitude and duration. Two physical regimes are clearly distinguished: a long pulse duration regime, in which reversal occurs by spin-transfer assisted thermal activation over an energy barrier, and a short time large pulse amplitude regime, in which the switching probability is determined by the spin angular momentum in the current pulse.
\end{abstract}

\maketitle

Magnetization dynamics of a nanomagnet in the presence of thermal noise is a topic of great fundamental interest and one of importance to magnetic information technologies. Of interest is the probability of a nanomagnet at finite temperature to reverse its direction of magnetization by thermal activation over an energy barrier. Spin-transfer torques \cite{Slonczewski1996,Berger1996,Katine2000,Sun2000} significantly affect thermally activated magnetization reversal, either enhancing or suppressing the reversal rate depending on the sign of the current \cite{Li2004,Apalkov2005}. Further, thermal fluctuations play a central role in spin-transfer device characteristics. A current greater than a critical current is needed to reverse the magnetization direction at zero temperature (i.e., in the absence of thermal noise) \cite{Sun2000}. However, spin-transfer driven reversal typically occurs for currents far less than this critical current because of thermal fluctuations. 

\begin{figure}[t]
\centering
\includegraphics[width=0.45\textwidth]{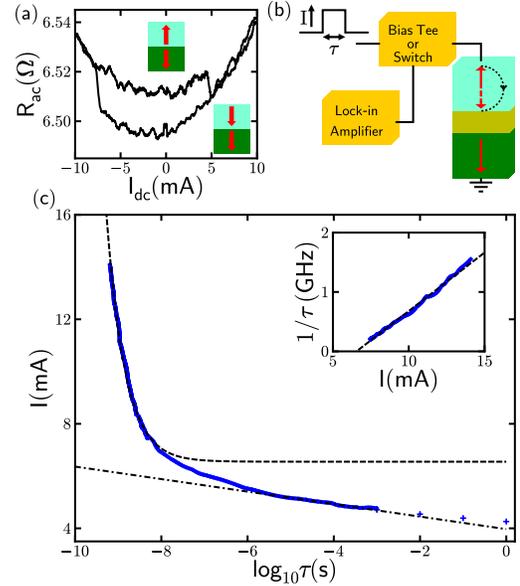}
\caption{(a) $R_{ac}$ vs. $I_{dc}$ of a $100 \;\unit{nm}$ device at room temperature in zero applied field. (b) Schematic of pulse experiments: a current pulse is applied to the nanopillar either via a bias tee or switch. (c) Pulse switching results for pulses of duration $0.3 \;\unit{ns}$ to $1\;\unit{s}$: pulse amplitude for $P=0.5$ vs. pulse duration. Inset: Switching rate ($1/\tau$) vs. pulse amplitude for short pulses showing $1/\tau \varpropto I$. The intercept ($1/\tau=0$) is $I_{c0}$. The dashed line in the main figure shows this behavior on a logarithmic scale. The dashed-dotted line is a fit to the long-time data using Eq.~\ref{eq:longBoundary}.}
\label{Figure1}
\end{figure}

Analytic models for finite temperature spin-torque dynamic have only been developed for uniaxial single domain nanomagnets, i.e. a nanomagnet with two energy minima, magnetization ``up'' and ``down'', separated by an anisotropy barrier  \cite{Li2004,Apalkov2005,He2007}. However, experiments have mainly explored soft thin film magnetic elements with a biaxial anisotropy, a shape anisotropy that leads to in-plane magnetization with a preferred axis in the plane \cite{Myers2002,Urazhdin2003,Koch2004}. The recent development of advanced nanopillar devices based on perpendicular magnetic anisotropy (PMA) materials \cite{Mangin2006,Mangin2009} now permit a direct comparison to analytic models. Further, most studies, including our own initial studies of pulse magnetization reversal \cite{Bedau2010}, examine the mean switching time, rather than the full probability distribution. Here we present results for spin-transfer switching of a uniaxial nanomagnet subject to current pulses from second to sub-nanosecond duration and measure the switching probability versus pulse duration.

Experiments were conducted at room temperature on spin-valve nanopillars that consist of two magnetic layers, a reference and a
free layer, both with a strong uniaxial anisotropy perpendicular to the plane of the layers. The free layer is a Co/Ni multilayer with a room temperature coercive field of about $0.1 \; \unit{T}$ and the reference layer consists of  Co/Pt and Co/Ni multilayers with a much larger coercive field ($\simeq 1$ T). The layer stack \cite{Bedau2010} is patterned into 
$50$ and $100 \;\unit{nm}$ square nanopillars with top and bottom electrical contacts such that current flows perpendicular to the plane of the layers. 22 junctions were studied and here we focus on the characteristics of two representative samples.

Figure~\ref{Figure1}(a) shows a measurement of the differential resistance versus dc current of a $100 \;\unit{nm}$ device in zero applied field. The high resistance branch corresponds to a state in which the free layer and reference layer magnetizations are antiparallel (AP) and the low resistance branch corresponds to the two layers magnetized in a parallel (P) configuration (MR=$0.3$\%).  A current of $4 \;\unit{mA}$ leads to single step switching of the device state from AP to P. Conversely, $-7 \;\unit{mA}$ changes the device state from P to AP.

To determine the switching probability as a function of time we apply current pulses of variable amplitude and duration, as illustrated in Fig.~\ref{Figure1}(b). The sample state was determined by measuring the device resistance using a small ($\le 300 \;\unit{\mu A}$) ac current and a lock-in amplifier before and after the pulse. The same pulse amplitude and duration is repeated $100$ to $10,000$ times to determine the switching probability. Since the state of the free layer is very stable in the absence of an applied field and dc current, switching only occurs through the application of a current pulse. We focus on AP to P switching events and have confirmed that similar behavior is observed for switching from the P to AP states.

Figure~\ref{Figure1}(c) shows the switching behavior from the AP to the P state
{\em over nine order of magnitude} in current pulse duration. The pulse amplitude $I$ corresponding
to the observation of switching for half of the events ($P=0.5$) is plotted versus 
pulse duration $\tau$ on a logarithmic scale. For short pulse durations ($\tau<10 \; \unit{ns}$) the pulse amplitude required to switch the nanomagnet increases dramatically. For long pulses the pulse amplitude depends weakly on pulse duration, varying logarithmically ($\tau>10 \; \unit{\mu s}$). 

The short and long time switching characteristics are thus quite distinct. The inset of Fig.~\ref{Figure1}(c) shows $1/\tau$ plotted versus pulse amplitude at short pulse durations. This boundary follows the form  $1/\tau = A(I - I_{c0})$, which is expected based on the conservation of angular momentum \cite{Sun2000}. The zero temperature critical current $I_{c0}$ in this expression reflects the portion of the spin-angular momentum that is needed to overcome the magnetization damping and the parameter $A$ characterizes the link between charge and spin-angular momentum transport. We find $I_{c0}=6.55 \; \unit{mA}$ and $A=2.0 \times 10^{11}\;\unit{C^{-1}}$ for this sample. The dashed line in the main part of Fig.~\ref{Figure1}(c) shows this same fit to the short time data on a logarithmic pulse duration scale, showing that this form describes the data up to pulse durations of $5 \;\unit{ns}$.

At long times the switching boundary was fit to the form \cite{Li2004,Apalkov2005}:
\begin{equation}
\label{eq:longBoundary}
I=I_{c0}\left(1- \frac{kT}{U_0} \ln(-\tau/(\tau_0 \ln(1-P))\right), 
\end{equation}
where $U_0$ is the energy barrier in the absence of the spin-current. This expression fits the data for $\tau>6 \; \unit{\mu s}$ well, as seen by the dashed-dotted line in Fig.~\ref{Figure1}(c). The slope of the curve gives the ratio $\xi=U_0/kT= 63$ and the intercept of this curve with $I_{c0}$ gives the Arrhenius prefactor, $\tau_0=24\; \unit{ps}$, which we discuss further below. 
Figure~\ref{Figure1}(c) also shows that at intermediate time scales, from $\sim 10\;\unit{ns}$ to $\sim 6 \;\unit{\mu s}$, the switching boundary follows neither the short or long time forms, i.e. this is a crossover regime.

\begin{figure}[t]
\includegraphics[width=0.5\textwidth]{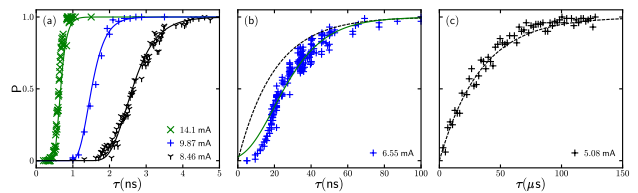}
\caption{Switching probability versus pulse duration in three regimes: 
(a) Short-time limit,  $I > I_{c0}=6.55 \;\unit{mA}$; (b)  Intermediate regime,  $I = I_{c0}$. (c) Long-time limit, $I < I_{c0}$. The solid lines are fits to Eq.~\ref{Shorttime} and the dashed curves are fits to Eq.~\ref{eq:longBoundary}.}
\label{Figure2}
\end{figure}

In order to understand the magnetization switching processes in more detail we studied the full switching probability distributions in the three regimes, i.e. at short, intermediate and long time scale (Fig.~\ref{Figure2}).  It is clear that the switching probability depends differently on pulse duration in these three cases. For short times the curves are sigmoidal and for long times the switching probability depends exponentially on pulse duration, $P=1-\exp(-\tau/\tau_A)$, with $\tau_A=\tau_0\exp(\xi(1-I/I_{c0}))$. A fit of this data gives $\tau_0=24 \;\unit{ps}$ and $\xi=63$, consistent with the fit to the long time data in Fig.~\ref{Figure1}(c). 

In the short pulse time limit the reversal process can no longer be considered thermal activation over an energy barrier. In this case, the current amplitude is larger than the critical current $I_{c0}$ and spin-torque quickly drives the magnetization reversal, starting from a thermally distributed initial state. The reversal time depends sensitively on the initial magnetization state and the distribution of initial magnetization states leads to a distribution of reversal times. This limit was considered theoretically in Ref.~\cite{He2007} for a uniaxial single domain nanomagnet. The following expressing for the switching probability was derived:
\begin{equation}
P=\exp \left(-{4 \xi} \exp \left(-\frac{2\tau (I/I_{c0}-1)}{\tau_D}\right) \right),
\label{Shorttime}
\end{equation}
where $\tau_D = 1/(\alpha \gamma \mu_0 H_k)$ is the characteristic time associated with the spin-transfer driven magnetization dynamics. $\alpha$ is the Gilbert damping, $\gamma$ is the gyromagnetic ratio and $H_k$ is the anisotropy field.  This expression fits the data well, as seen in Fig.~\ref{Figure2}(a), with $\xi=63$ and $\tau_D=260 \;\unit{ps}$.

At intermediate times (Fig.~\ref{Figure2}(b)), the switching probability has a distinct form, it is neither the sigmoidal form measured at short times (the solid curve in Fig.~\ref{Figure2}(b)) nor the exponential form (the dashed curve in Fig.~\ref{Figure2}(b)) observed at longer time scales. On such time scales thermal fluctuations during magnetization reversal and spin-torque driven processes are both important--and cannot be separated.

\begin{figure}[t]
\includegraphics[width=0.4\textwidth]{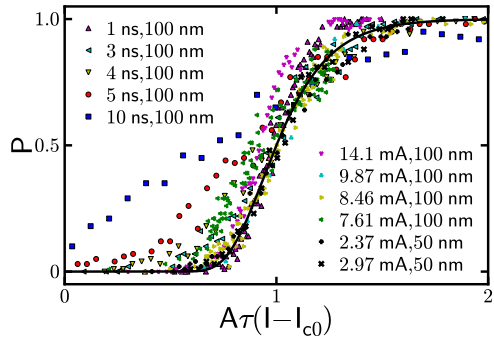}
\caption{Switching probability as a function of the scaled angular momentum ,$A\tau (I - I_{c0})$, in the short time regime for both $50$ and $100 \;\unit{nm}$ devices. Both data taken with variable pulse duration and variable pulse amplitude are shown. For $\tau \lesssim 5 \;\unit{ns}$ $P$ only depends on the net spin angular momentum in the pulse.}
\label{Figure3} 
\end{figure}

An interesting prediction of the model for short time switching  (Eq.~\ref{Shorttime}) is that the switching probability only depends on the total spin angular momentum in the pulse, $p\hbar I \tau /e$, where $p$ is the spin-polarization of the current  \cite{He2007}. We test this model and more generally whether the spin angular momentum is the relevant experimental variable by plotting the switching probability in Fig.~\ref{Figure2}(a) versus the scaled angular momentum, $A \tau (I-I_{c0})$ (Fig.~\ref{Figure3}). Included in this plot are additional measurements of the switching probability in which we varied the pulse amplitude for a constant pulse duration. Figure~\ref{Figure3} shows that there is a good scaling of the data, particularly for short pulse durations. Deviations from the scaling form (Eq.~\ref{Shorttime}) occur for pulse durations of $5 \; \unit{ns}$ and greater. This indicates the time scales on which thermal fluctuations influence the switching process during the current pulse. Pulse studies conducted on a $50\;\unit{nm}$ device are also shown in Fig.~\ref{Figure3}. For this sample $I_{c0}=1.04 \; \unit{mA}$ and $A=1.5 \times 10^{12}\; \unit{C^{-1}}$. Data in this case, with pulse amplitudes $I>2I_{c0}$, closely follow the scaling curve.

The single domain model assumes an energy density:
$E=- \mu_0 M_sH_km_z^2/2$, where $m_z$ is the normalized magnetization in the z-direction. 
The energy barrier is $U_0= \mu_0 M_sH_k V/ 2 $. $\mu_0 H_k$ of our nanomagnet is $0.25 \;\unit{T}$ giving an energy barrier of $\xi=360$ at room temperature, six times larger than the energy barrier found from the data in Fig.~\ref{Figure1}(c) and Fig.~\ref{Figure2}(a),(c). This implies that the nanomagnet is not reversing as a single magnetic domain. Further, the attempt time in a single domain model is given by $\tau_0 = \sqrt{\pi}(1+ \alpha^2)/(2\gamma \mu_0 H_k \alpha\sqrt{\xi})$ \cite{Brown1963,Coffey1998}. With $\xi = 63$ and $\alpha=0.01$ we find $\tau_0 = 1.6\;\unit{ns} $ and with $\alpha = 0.1$, $\tau_0 = 162 \; \unit{ps} $. In either case, this time is significantly larger than the $24\;\unit{ps}$ determined in our experiment. While the $\tau_0$ we find is small compared to the macrospin model, it is within the range of values determined in other experimental studies. Krause {\it et al.} \cite{Krause2009} report even shorter $\tau_0$, from $10^{-16}$ to $10^{-13} \;\unit{s},$ for nanostructures studied by scanning tunneling microscopy. In the short time limit the single domain model predicts that $\tau_D=(\alpha \gamma \mu_0 H_k)^{-1}$. Taking $\alpha=0.01$ to $0.1$ gives $\tau_D=1.4$ to $14\; \unit{ns}$, far larger than $\tau_D$ deduced from the data in Fig.~\ref{Figure2}(a), $\tau_D=0.26 \;\unit{ns}$. These comparisons show that the microscopic time scales are shorter than those predicted in the single domain model.

Our results demonstrate that there are two distinct spin torque switching processes, spin-transfer assisted thermal activation at long times and one dominated by angular momentum conservation at short times. The measured switching probability distributions at both short and long times are in good agreement with a single domain model that includes spin-transfer torques and thermal fluctuations \cite{Li2004,Apalkov2005,He2007}. However, parameters such as the energy barrier deduced from fits to these models are far less than that expected. This suggests that the switching occurs by rotation of small part of the nanomagnet, whose dynamics is nonetheless captured by a macrospin model, as was found in field driven reversal of nanostructures with PMA \cite{Thomson2006}. 

The research at NYU was supported by USARO Grant No. W911NF0710643, NSF-DMR Grant Nos. 0706322 and 1006575.

\end{document}